# WAKE FIELDS IN A RECTANGULAR DIELECTRIC-LINED ACCELERATING STRUCTURE WITH TRANSVERSAL ISOTROPIC LOADING

I. L. Sheinman[*], Yu. S. Sheinman,
[*]ishejnman@yandex.ru
Saint-Petersburg Electro-Technical University «LETI», Saint-Petersburg, Russia

*Abstract*

Dielectric lined waveguides are under extensive study as accelerating structures that can be excited by electron beams. Rectangular dielectric structures are used both in proof of principle experiments for new accelerating schemes and for studying the electronic properties of the structure loading material. Analysis of Cherenkov radiation generated by high current relativistic electron bunch passing through a rectangular waveguide with transversal isotropic dielectric loading has been carried out. Some of the materials used for the waveguide loading of accelerating structures (sapphire, ceramic films) possess significant anisotropic properties. In turn, it can influence excitation parameters of the wakefields generated by an electron beam. General solutions for the fields generated by a relativistic electron beam propagating in a rectangular dielectric waveguide have been derived using the orthogonal eigenmode decomposition method for the transverse operators of the Helmholtz equation. The analytical expression for the combined Cherenkov and Coulomb fields in terms of a superposition of LSM and LSE-modes of rectangular waveguide with transversal isotropic dielectric loading has been obtained. Numerical modelling of the longitudinal and transverse (deflecting) wakefields has been carried out as well. It is shown that the dielectric anisotropy causes frequency shift in comparison to the dielectric-lined waveguide with the isotropic dielectric loading.

## INTRODUCTION

Physics of particle accelerator now is on the edge of traditional and new accelerating methods. One of the promising directions is development of linear colliders with high acceleration rate on the base of wakefield accelerating structures. Wakefield waveguide structures can contain plasma or dielectric loading excited by laser, RF source or high current charged beam. Unlike plasma structures, dielectric filled waveguides with vacuum channel provide collisionless transport of the beam [1, 2]. The Cherenkov accelerating structure is a dielectric waveguide with an axial vacuum channel for beam passing covered by conductive sleeve.

A high current electron beam or a high power RF source can excite these structures. The high current short generating bunch (usually called driver) with low energy excites Vavilov-Cherenkov wake field. Generated longitudinal field accelerates a low intensive but higher energy bunch (witness). The witness is placed to a distance behind the driving bunch corresponds to an accelerating phase of the wake field.

Dielectric wakefield structures provide both high acceleration rate and ensure the control over the frequency spectrum of the structure by introducing additional ferroelectric layers [3] as well as a possibility using of perspective materials with unique properties like diamond and sapphire [4].

As a rule, the cylindrical geometry proposed for structures with dielectric loading is essential for attaining the highest accelerating gradients as well as for obtaining the maximal possible shunt impedance of the structure [2, 4]. Analytic mode analysis of such accelerating structures for the longitudinal and transverse electric field components was developed in a number of publications (see, for example, [5]). At the same time, structures with a rectangular cross section and dielectric loading were also considered in some cases [6–15] in view of technological difficulties in preparing cylindrical structures with stringent requirements to tolerances for geometrical parameters and uniformity of the permittivity of the filling along the structures [3], as well as their possible application for generating a sheet electron beam.

Advantage in usage of this geometry is simplification of manufacturing techniques. Such structures (along with cylindrical structures) for generating electromagnetic radiation and producing wakefield acceleration in the frequency range 0.5–1.0 THz are considered [4]. In THz range, the planar geometry can be preferable because of difficulties of precise cylindrical structure manufacture.

Rectangular structures can be used for test experiments in analysis of new accelerating systems [15] and for studying the properties of materials effective for producing high acceleration gradients of the structure (diamond, sapphire) [4].

Theoretical analysis of dielectric accelerating structures of rectangular geometry has been carried out in a number of publications [6–9, 13–15]. To determine the amplitudes of individual Cherenkov radiation modes excited in a rectangular waveguide with a dielectric loading, the impedance matching technique was used earlier [7–9]. When such formalism was used instead of direct solution of the nonhomogeneous system of Maxwell equations (which is a standard analytic approach in analysis of wake fields in cylindrical structures [5]), the amplitudes of wake fields had to be expressed in terms of the shunt impedance (or integrated loss factor) for each mode of the structure. Such an approach involves certain approximations, while direct solution of the non homogeneous system of Maxwell's equations without indirect constructions is always preferable for analyzing the problems of generation in waveguide structures.

A method of the first order transverse operator as applied to waveguide problems was worked out in [10–12]. In [13], the generalized orthogonality relation between the LSM and LSE modes was derived. However, the bilinear form introduced in [13] is not the scalar product in the $L_2$ space, which requires a substantiation of the possibility of application of this relation for describing orthogonality between the components of the electric and magnetic field vectors. In [14, 15], analysis was performed on the basis of the construction of second order differential equations for transverse field components (based on direct solution of the Maxwell equations) of the two channel rectangular structure with dielectric filling; this structure was developed for increasing the energy conversion factor from the leading beam to the beam being accelerated. In [16] a strict theory of beam excitation of rectangular waveguide structure with isotropic dielectric loading was developed. We used similar method for our case but with some improvements in finding of longitudinal wake fields.

## THEORETICAL ANALYSES OF RECTANGULAR WAVEGUIDE EXCITATION

Let us consider a rectangular waveguide with a symmetric filling in the form of dielectric transversal isotropic layers parallel to the $x$ axis and with a vacuum channel at the centre (Fig. 1).

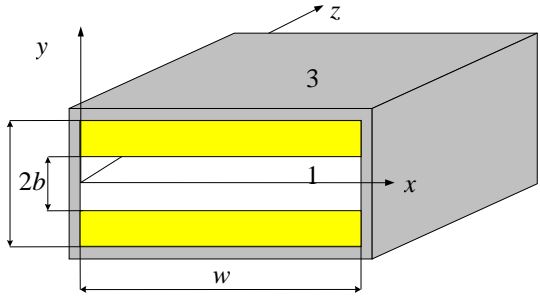

Figure 1: Rectangular waveguide.

In this case, the filling in the direction of the $y$ axis is inhomogeneous, and the permittivity and permeability tensors are functions of y: $\varepsilon = \varepsilon(y)$ and $\mu = \mu(y)$.

$$\varepsilon = \begin{pmatrix} \varepsilon_{\parallel}(y) & 0 & 0 \\ 0 & \varepsilon_{\perp}(y) & 0 \\ 0 & 0 & \varepsilon_{\parallel}(y) \end{pmatrix},$$

$$\mu = \begin{pmatrix} \mu_{\parallel}(y) & 0 & 0 \\ 0 & \mu_{\perp}(y) & 0 \\ 0 & 0 & \mu_{\parallel}(y) \end{pmatrix}.$$

Let us transform initial Maxwell equations combined with material relations for this case.

Maxwell equations can be transformed to equations for normal to dielectric layer electric and magnetic field components in isotropic [15-16] and anisotropic [17,18] cases.

$$\frac{\partial^2 E_y}{\partial \zeta^2} + T_E E_y = \frac{-e}{\varepsilon_0 \left(1 - \varepsilon_\perp \mu_\parallel \beta^2\right)} \frac{\partial}{\partial y}\left(\frac{n}{\varepsilon_\parallel}\right), \quad (1)$$

$$\frac{\partial^2 H_y}{\partial \zeta^2} + T_H H_y = \frac{-ev}{1 - \varepsilon_\parallel \mu_\perp \beta^2} \left(\frac{\partial n}{\partial x}\right), \quad (2)$$

where

$$T_E E_y = \frac{1}{\left(1 - \varepsilon_\perp \mu_\parallel \beta^2\right)} \left[\frac{\partial^2 E_y}{\partial x^2} + \frac{\partial}{\partial y}\left(\frac{1}{\varepsilon_\parallel} \frac{\partial \left[\varepsilon_\perp E_y\right]}{\partial y}\right)\right],$$

$$T_H H_y = \frac{1}{\left(1 - \varepsilon_\parallel \mu_\perp \beta^2\right)} \left[\frac{\partial^2 H_y}{\partial x^2} + \frac{\partial}{\partial y}\left(\frac{1}{\mu_\parallel} \frac{\partial \left[\mu_\perp H_y\right]}{\partial y}\right)\right].$$

Equation (1) corresponds to LSM waveguide modes, (2) corresponds to LSE modes. Equations give biorthogonality of the eigenfunctions and similarity of the operator to a self adjoint operator [15-16].

Normalized eigenfunctions $Y_{E_y}(y)$ and $Y_{H_y}(y)$ break up to system of symmetric and antisymmetric modes concerning a wave guide axis.

For vacuum channel
$$k_{y1} = \sqrt{k_x^2 + \lambda_E \left(1 - \beta^2\right)},$$

For the dielectric
$$k_{y2E} = \sqrt{\frac{\varepsilon_{\parallel 2}}{\varepsilon_{\perp 2}}\left[\left(\varepsilon_{\perp 2}\mu_{\parallel 2}\beta^2 - 1\right)\lambda_E - k_x^2\right]},$$

$$k_{y2H} = \sqrt{\frac{\mu_{\parallel 2}}{\mu_{\perp 2}}\left[\left(\varepsilon_{\parallel 2}\mu_{\perp 2}\beta^2 - 1\right)\lambda_H - k_x^2\right]}.$$

The dispersion equation for the symmetrical concerning the waveguide axes LSM modes
$$\varepsilon_{\parallel 2} k_{y1} \tanh\left(k_{y1} b\right) - \varepsilon_1 k_{y2E} \tan\left(k_{y2E}(c-b)\right) = 0.$$

To system of antisymmetric LSM modes there corresponds the dispersion equation
$$\varepsilon_{\parallel 2} k_{y1} \coth\left(k_{y1} b\right) - \varepsilon_1 k_{y2E} \tan\left(k_{y2E}(c-b)\right) = 0.$$

Then

$$Y_{E_y}(y) = A_E \begin{cases} \dfrac{1}{\varepsilon_{\perp 2}}\cos(k_{y2E}(c-y)), \\[4pt] \begin{bmatrix}\cosh\\ \sinh\end{bmatrix}(k_{y1}y)\dfrac{\cos(k_{y2E}(c-b))}{\varepsilon_1 \begin{bmatrix}\cosh\\ \sinh\end{bmatrix}(k_{y1}b)}, \\[4pt] \dfrac{1}{\varepsilon_{\perp 2}}\cos(k_{y2E}(c+y)), \end{cases}$$

$$Y_{D_y}(y) = A_E \begin{cases} \cos(k_{y2E}(c-y)), \\[4pt] \begin{bmatrix}\cosh\\ \sinh\end{bmatrix}(k_{y1}y)\dfrac{\cos(k_{y2E}(c-b))}{\begin{bmatrix}\cosh\\ \sinh\end{bmatrix}(k_{y1}b)}, \\[4pt] \cos(k_{y2E}(c+y)), \end{cases}$$

$$Y_{Ed}(y) = A_E \begin{cases} \dfrac{k_{y2E}}{\varepsilon_{\|2}}\sin(k_{y2}(c-y)), \\ k_{y1}\begin{bmatrix}\sinh\\\cosh\end{bmatrix}(k_{y1}y)\dfrac{\cos(k_{y2}(c-b))}{\varepsilon_{\|1}\begin{bmatrix}\cosh\\\sinh\end{bmatrix}(k_{y1}b)}, \\ -\dfrac{k_{y2}}{\varepsilon_{\|2}}\sin(k_{y2}(c+y)), \end{cases}$$

Here in denotations $\begin{bmatrix}\cdot\\\cdot\end{bmatrix}$ the upper choice corresponds to symmetrical mode of eigenfunction, and the lower choice corresponds to antisymmetrical one.

Normalizing coefficient is

$$A_E = \sqrt{\dfrac{2}{w}\left((1-\varepsilon_1\mu_1\beta^2)\dfrac{\cos^2(k_{y2E}(c-b))}{\varepsilon_1\left(\begin{bmatrix}\cosh\\\sinh\end{bmatrix}(k_{y1}b)\right)^2}\left(\dfrac{\sh(2k_{y1}b)}{2k_{y1}}[\pm]b\right)+\right.}$$
$$\overline{\left.+\dfrac{(1-\varepsilon_{\perp 2}\mu_{\|2}\beta^2)}{\varepsilon_{\perp 2}}\left(c-b+\dfrac{\sin(2k_{y2E}(c-b))}{2k_{y2E}}\right)\right)^{-\tfrac{1}{2}}}.$$

The dispersion equation for the symmetric LSE modes:
$$\mu_{\|2}k_{y1}\tanh(k_{y1}b)+\mu_1 k_{y2H}\cot(k_{y2H}(c-b))=0.$$

To system of antisymmetric LSE modes there corresponds the dispersion equation
$$\mu_1 k_{y2H}\tanh(k_{y1}b)+\mu_{\|2}k_{y1}\tan(k_{y2H}(c-b))=0.$$

Then

$$Y_{H_y}(y) = A_H \begin{cases} \dfrac{1}{\mu_{\perp 2}}\sin(k_{y2H}(c-y)), \\ \begin{bmatrix}\ch\\\sh\end{bmatrix}(k_{y1}y)\dfrac{\sin(k_{y2H}(c-b))}{\mu_1\begin{bmatrix}\ch\\\sh\end{bmatrix}(k_{y1}b)}, \\ \dfrac{1}{\mu_{\perp 2}}\sin(k_{y2H}(c+y)), \end{cases}$$

$$Y_{B_y}(y) = A_H \begin{cases} \sin(k_{y2H}(c-y)), \\ \begin{bmatrix}\ch\\\sh\end{bmatrix}(k_{y1}y)\dfrac{\sin(k_{y2H}(c-b))}{\begin{bmatrix}\ch\\\sh\end{bmatrix}(k_{y1}b)}, \\ \sin(k_{y2H}(c+y)), \end{cases}$$

$$Y_{Hd}(y) = A_H \begin{cases} -\dfrac{k_{y2H}}{\mu_{\|2}}\cos(k_{y2H}(c-y)), \\ k_{y1}\begin{bmatrix}\sh\\\ch\end{bmatrix}(k_{y1}y)\dfrac{\sin(k_{y2H}(c-b))}{\mu_1\begin{bmatrix}\ch\\\sh\end{bmatrix}(k_{y1}b)}, \\ \dfrac{k_{y2H}}{\mu_{\|2}}\cos(k_{y2H}(c+y)), \end{cases}$$

Normalizing coefficient is

$$A_H = \sqrt{\dfrac{2}{w}\left((1-\varepsilon_1\mu_1\beta^2)\dfrac{\sin^2(k_{y2E}(c-b))}{\mu_1\left(\begin{bmatrix}\cosh\\\sinh\end{bmatrix}(k_{y1}b)\right)^2}\left(\dfrac{\sh(2k_{y1}b)}{2k_{y1}}[\pm]b\right)+\right.}$$
$$\overline{\left.+\dfrac{(1-\varepsilon_{\|2}\mu_{\perp 2}\beta^2)}{\mu_{\perp 2}}\left(c-b-\dfrac{\sin(2k_{y2E}(c-b))}{2k_{y2E}}\right)\right)^{-\tfrac{1}{2}}}.$$

Solutions for all field components are:

$$E_y = \sum_{n,m}\dfrac{\psi(x,x_0)}{\varepsilon_0}Y_{E_y}(y)Y_{Ed}(y_0)S_E(\zeta,\zeta_0),$$

$$E_z = -\sum_{n,m}^{\infty}\dfrac{\psi(x,x_0)}{\varepsilon_0}\left[\dfrac{G_E(\zeta,\zeta_0)Y_{Ed}(y)Y_{Ed}(y_0)}{(k_{xn}^2+\lambda_E)}+\right.$$
$$\left.+\dfrac{k_{xn}^2\beta^2 G_H(\zeta,\zeta_0)Y_{B_y}(y)Y_B(y_0)}{(k_{xn}^2+\lambda_H)}\right],$$

$$E_x = \sum_{n,m}^{\infty}\dfrac{\psi'(x,x_0)}{\varepsilon_0}\left[\dfrac{S_E(\zeta,\zeta_0)Y_{Ed}(y)Y_{Ed}(y_0)}{(k_{xn}^2+\lambda_E)}-\right.$$
$$\left.-\dfrac{\lambda_H S_H(\zeta,\zeta_0)\beta^2}{(k_{xn}^2+\lambda_H)}Y_{B_y}(y)Y_B(y_0)\right],$$

$$H_y = -v\sum_{n,m}\psi'(x,x_0)Y_{H_y}(y)Y_B(y_0)S_H(\zeta,\zeta_0),$$

$$H_z = v\sum_{n,m}\psi'(x,x_0)\left[\dfrac{G_H(\zeta,\zeta_0)Y_{Hd}(y)Y_B(y_0)}{(k_{xn}^2+\lambda_H)}+\right.$$
$$\left.+\dfrac{G_E(\zeta,\zeta_0)Y_{D_y}(y)Y_{Ed}(y_0)}{(k_{xn}^2+\lambda_E)}\right],$$

$$H_x = v\sum_{n,m}\psi(x,x_0)\left[\dfrac{k_{xn}^2 S_H(\zeta,\zeta_0)Y_{Hd}(y)Y_B(y_0)}{(k_{xn}^2+\lambda_H)}-\right.$$
$$\left.-\dfrac{\lambda_E S_E(\zeta,\zeta_0)Y_{D_y n,m}(y)Y_{Ed}(y_0)}{(k_{xn}^2+\lambda_E)}\right].$$

Here we used the following designations:
$$\psi(x,x_0) = q\sin(k_{xn}x)\sin(k_{xn}x_0),$$
$$\psi'(x,x_0) = qk_{xn}\cos(k_{xn}x)\sin(k_{xn}x_0),$$
$$G_{E,H}(\zeta,\zeta_0) = \begin{cases}\cos\left(\sqrt{\lambda_{E,H}}(\zeta-\zeta_0)\right)\theta(\zeta_0-\zeta), \lambda_{E,H}\geq 0;\\ \dfrac{\sign(\zeta-\zeta_0)}{2}e^{-\sqrt{|\lambda_{E,H}|}|\zeta-\zeta_0|}, \lambda_{E,H}<0,\end{cases}$$

$$S_{E,H}(\zeta,\zeta_0) = \begin{cases} -\dfrac{\sin\left(\sqrt{\lambda_{E,H}}(\zeta-\zeta_0)\right)}{\sqrt{\lambda_{E,H}}}\theta(\zeta_0-\zeta), & \lambda_{E,H} \geq 0; \\ \dfrac{\text{sign}(\zeta-\zeta_0)}{2\sqrt{|\lambda_{E,H}|}} e^{-\sqrt{|\lambda_{E,H}|}|\zeta-\zeta_0|}, & \lambda_{E,H} < 0, \end{cases}$$

$\theta(\zeta)$ is the Heaviside function.

The transversal forces operating on electrons in a waveguide can be found with use of a formula of Lorentz:

$$\frac{F_x}{-e} = \sum_{n,m}^{\infty} \frac{\psi'(x,x_0)}{\varepsilon_0}\left[\frac{S_E(\zeta,\zeta_0)Y_{Ed}(y)Y_{Ed}(y_0)}{(k_{xn}^2+\lambda_E)} + \frac{S_H(\zeta,\zeta_0)k_{xn}^2\beta^2 Y_{B_y}(y)Y_B(y_0)}{(k_{xn}^2+\lambda_H)}\right],$$

$$\frac{F_y}{-e} = \sum_{n,m} \frac{\psi(x,x_0)}{\varepsilon_0}\left[\frac{k_{xn}^2\beta^2\mu_\| S_H(\zeta,\zeta_0)Y_{Hd}(y)Y_B(y_0)}{(k_{xn}^2+\lambda_H)} + \left(S_E(\zeta,\zeta_0)\left(\frac{1}{\varepsilon_\perp} - \frac{\mu_\|\lambda_E\beta^2}{k_{xn}^2+\lambda_E}\right)\right)Y_{D_y}(y)Y_{Ed}(y_0)\right].$$

Transverse fields can be used to calculate self-consistent beam dynamics in the rectangular dielectric wakefield waveguides.

## CALCULATION RESULTS

The expressions derived above were used for analyzing the wakefields generated by a Gaussian relativistic electron bunch with parameters of the Argonne Wakefield Accelerator in the sapphire-based rectangular accelerating structure [4]: $w = 11$ mm, $b = 1.5$ mm, $c = 2.39$ mm, $\varepsilon_{2\perp} = 11.5$, $\varepsilon_{2\|} = 9.4$ (Fig. 1), which corresponds to a frequency of 25.0 GHz of the accelerating LM mode of the structure. For comparison a waveguide with isotropic dielectric filling with the same parameters but $\varepsilon_2 = 11.5$ corresponds to the base frequency of 23.25 GHz, $\varepsilon_2 = 10.45$ corresponds to the base frequency of 24.23 GHz, $\varepsilon_2 = 9.4$ corresponds to the base frequency of 25.36 GHz.

As a source of Cherenkov radiation, a generator electron bunch with a Gaussian charge distribution and energy $W = 15$ MeV, charge $q = 100$ nC and bunch length $\sigma_z = 1.5$ mm was considered. The dependence of the longitudinal electric field component $E_z$ produced by the bunch on the distance $\xi = z - vt$ behind it is shown in Fig. 2 (the bunch is located at point $x_0 = w/2$, $y_0 = 0$, $\xi_0 = 8$ cm; the coordinates of the observation point are $x = w/2$, $y = 0$, $\xi = z - vt$); the high accelerating gradient (exceeding 100 MV/m) of wake radiation behind the bunch is worth noting. Solid line corresponds to anisotropic sapphire with $\varepsilon_{2\perp} = 11.5$, $\varepsilon_{2\|} = 9.4$, dashed line corresponds to isotropic filling with $\varepsilon_2 = 10.45$. It is visible that sapphire anisotropy leads to shift of a frequency range of the waveguide essential to wakefield acceleration, but with a little influence on a wake field amplitudes.

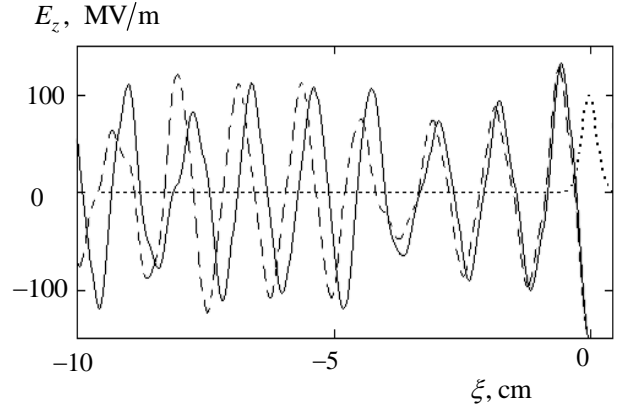

Figure 2: Longitudinal wake field.

Let us consider wake fields in the sapphire-based rectangular sub-THz accelerating structure generated by a Gaussian relativistic electron bunch. The accelerating structure has the following parameters: $w = 2.5$ mm, $b = 1.0$ mm, $c = 1.04$ mm, $\varepsilon_{2\perp} = 11.5$, $\varepsilon_{2\|} = 9.4$, base frequency is 300 GHz. A generator electron bunch has a Gaussian charge distribution and energy $W = 75$ MeV, charge $q = 10$ nC and bunch length $\sigma_z = 0.1$ mm. The bunch is located at point $x_0 = w/2$, $y_0 = 0$, $\xi_0 = 0$ cm; the coordinates of the observation point are $x = w/2$, $y = 0$, $\xi = z - vt$.

The dependence of the longitudinal electric field component $E_z$ produced by the bunch on the distance $\xi = z - vt$ behind it is shown in Fig. 3. The high accelerating gradient (exceeding 100 MV/m) combined with one-mode regime of wake radiation behind the bunch is worth noting.

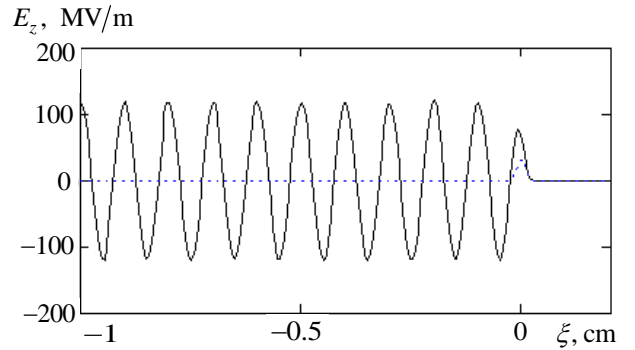

Figure 3: Dependency on the distance of the longitudinal wake field in the sub-THz rectangular accelerating structure.

Dependences of electric (Fig. 4) and magnetic (Fig. 5) fields and transversal forces (Fig. 6) on coordinates were simulated. Simulation was done for z-coordinate corresponding to the first maximum of the longitudinal electric field followed the driver bunch.

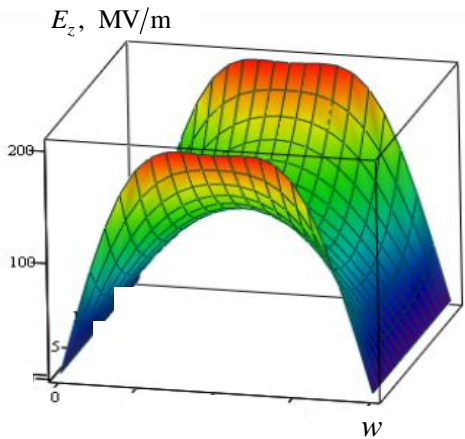

Figure 4: Spatial distribution of the electric intensity.

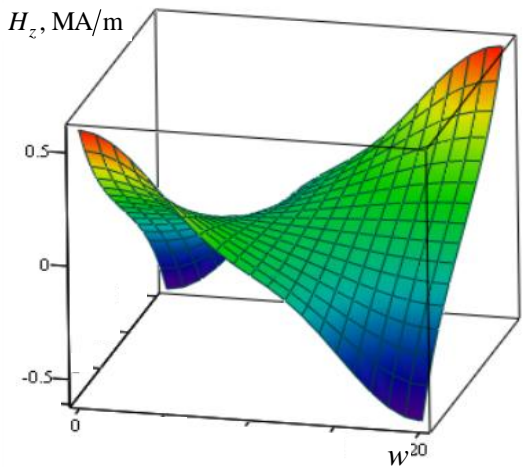

Figure 5: Spatial distribution of the magnetic strength.

Electric field strength has maximum near the borders of dielectric loading in the centre of waveguide along x-axis. It is twice higher than the field strength near the waveguide axis of symmetry. In the centre of the waveguide, accelerating gradients higher than 100 MV/m were obtained. Longitudinal magnetic field strength along the waveguide axis of symmetry is rather small – about zero, but dramatically rises as the corners of dielectric filling approached.

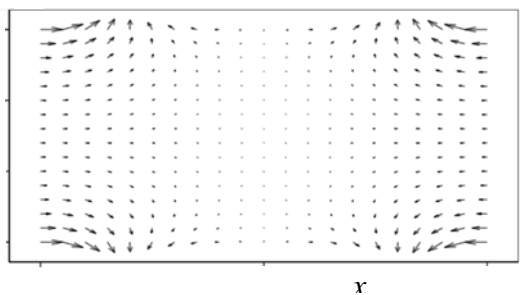

Figure 6: Transversal forces.

Transversal Lorentz forces are rather small at the center of the waveguide. Force increases as the dielectric edge is approached.

## SUMMARY

We have proposed a method for calculating wake fields of Cherenkov radiation in a rectangular accelerating structure with transversal-isotropic dielectric filling. Using this method, we have analyzed the sapphire based dielectric structure with a rectangular cross section, in which accelerating gradients higher than 100 MV/m can be attained.

## REFERENCES

[1] W. Gai, P. Schoessow, B. Cole, R. Konecny, et al., Phys. Rev. Lett. 61, 2756 (1988).
[2] W. Gai, AIP Conf. Proc. 1086, 3 (2009).
[3] A. M. Al'tmark, A. D. Kanareikin, and I. L. Sheinman, Tech. Phys. 50, 87 (2005).
[4] A. D. Kanareykin, J. Phys.: Conf. Ser. 236, 012032 (2010).
[5] A. D. Kanareikin and I. L. Sheinman, Tech. Phys. Lett. 33, 344 (2007).
[6] L. Xiao, W. Gai, and X. Sun, Phys. Rev. E 65, 1 (2001).
[7] C. Jing, W. Liu, W. Gai, L. Xiao, and T. Wong, Phys.Rev. E 68, 016502 (2003).
[8] A. Tremaine, J. Rosenzweig, P. Schoessow, and W. Gai, Phys. Rev. E 56, 7204 (1997).
[9] A. D. Bresler, G. H. Joshi, and N. Marcuvitz, J. Appl. Phys. 29, 794 (1958).
[10] A. Rowland and J. Sammut, J. Opt. Soc. Am. 72, 1335 (1982).
[11] J. W. Tao, J. Atechian, P. Ratovondrahanta, and H. Baudrand, Proc. IEE 137 (Part H), 311 (1990).
[12] S. Y. Park, C. Wang, and J. L. Hirshfield, AIP Conf. Proc. 647, 527 (2002).
[13] C. Wang and J. L. Hirshfield, Phys. Rev. ST Accel. Beams 9, 031301 (2006).
[14] G. V. Sotnikov, I. N. Onishchenko, J. L. Hirshfield, and T. C. Marshal, Probl. At. Nauki Tekhnol. Ser.: Yad._Fiz.Issled., No. 3 (49), 148 (2008).
[15] S. S. Baturin, I. L. Sheinman, A. M. Altmark, A. D. Kanareikin, Tech. Phys. 57, 5, 683 (2012)
[16] S. S. Baturin, I. L. Sheinman, A. M. Altmark, and A. D. Kanareykin. Transverse Operator Method for Wakefields in a Rectangular Dielectric Loaded Accelerating Structure. Phys. Rev. ST Accel. Beams 16, 051302 (2013)
[17] I. Sheinman, S. Baturin, A. Kanareykin. "Analysis of a Rectangular Dielectric-lined Accelerating Structure with an Anisotropic Loading" IPAC'12, New Orlean, USA, pp. 2769-2771. (2012)
[18] I. Sheinman, Yu. Sheinman. "Wake Field Components in a Rectangular Accelerating Structure with Dielectric Anisotropic Loading". RUPAC-2016, TUPSA043, 21-25 November 2016, Saint-Petersburg, Russia